\documentclass {revtex4}
\usepackage{amssymb}
\usepackage{epsfig}
\usepackage{graphicx}
\usepackage{dcolumn}
\usepackage{bm}

\begin{document}

\title{Twenty five years after KLS: A celebration of non-equilibrium statistical
mechanics }
\author{R. K. P. Zia}
\affiliation{Department of Physics, Virginia Polytechnic Institute and State
University, Blacksburg, VA 24061-0435, USA}
\date{November 11, 2009}

\begin{abstract}
When Lenz proposed a simple model for phase transitions in magnetism, he
couldn't have imagined that the ``Ising model" was to become a jewel in
field of equilibrium statistical mechanics. Its role spans the spectrum,
from a good pedagogical example to a universality class in critical
phenomena. A quarter century ago, Katz, Lebowitz and Spohn found a similar
treasure. By introducing a seemingly trivial modification to the Ising
lattice gas, they took it into the vast realms of non-equilibrium
statistical mechanics. An abundant variety of unexpected behavior emerged
and caught many of us by surprise. We present a brief review of some of the
new insights garnered and some of the outstanding puzzles, as well as
speculate on the model's role in the future of non-equilibrium statistical
physics.

\end{abstract}

\maketitle

\section{Introduction}

\label{intro}

Over a century ago, Boltzmann and Gibbs laid the foundations for a
comprehensive treatment of all systems in thermal equilibrium. By contrast,
our understanding of non-equilibrium statistical mechanics (NESM) is quite
primitive. To date, an overarching principle remains elusive, even for
time-independent (steady) states. Yet, such systems are ubiquitous,
encompassing all biological systems, for example. Thus, understanding
``physics far from equilibrium'' is recognized as one of the greatest
challenges of current condensed matter and materials physics \cite{CMMP2010}%
, with significant implications for technological advances in the sciences
and engineering. Faced with such grand vistas, one approach is to focus on
simple model systems, with the goal of identifying essential characteristics
of NESM that defy our equilibrium-trained expectations and intuitions. This
paper will be devoted to a seminal effort in this direction.

Twenty five years - and 50 Statistical Mechanics Meetings - ago, Katz,
Lebowitz and Spohn (KLS) introduced a simple model\cite{KLS}, motivated
partly by the physics of fast ionic conductors under the influence of an
external DC field \cite{FIC}. It involves a seemingly trivial modification
of the two-dimensional Lenz-Ising model\cite{Ising}, namely, an interacting
lattice gas with {\em biased} hopping along one of the axes. A brief review
of this model is the main theme of this article, dedicated to the
celebration of the 100$^{th}$ Statistical Mechanics Meeting. We will recount
the many surprises it provided, the numerous variations it spawned, the
aspects that are now understood, the puzzles that remain outstanding, as
well as some of the general insights on NESM it offered. A more detailed
review of the ``first dozen years of KLS'' can be found in \cite{DL17}.

The KLS and the Ising models share many noteworthy features. Both represent
the barest of essentials for a system with many interacting degrees of
freedom to display non-trivial behavior. Each is motivated by physical
systems and their interesting properties, e.g., phase transitions. In the
20's, the Ising model was ``scorned or ignored''\cite{SGB67} as a simple
mathematical toy of theorists. When Lenz proposed the model, Ising was able
to solve it only in one dimension and the resultant lack of a phase
transition must have disappointed theorists at the time. With the
contributions of many, from Onsager\cite{Onsager} to Wilson and Fisher\cite
{WF}, it became part of well-established text-book material. Not only is it
one of the most celebrated models within and beyond theoretical physics, it
has been realized physically in several systems\cite{RealIsing}.
Fortunately, the KLS model enjoyed more respect in its first decade than the
Ising model did in the 30's. Unfortunately, a physical realization has yet
to be found. Furthermore, it is so challenging that very little is known
analytically, even in one dimension. At the ``most fundamental'' level, its
stationary distribution is not known, in stark contrast to the explicit $%
\exp \left[ \beta J\sum ss^{\prime }\right] $ in the Ising model in
equilibrium. Consequently, most of its macroscopic properties are beyond our
analytic abilities and its collective behavior, as discovered through
computer simulations, continues to confound our intuitive ideas. In the
remainder of this article, we briefly look back on these 25 years and look
forward to much more progress on the KLS model, as well as non-equilibrium
statistical mechanics in general.

\section{The driven lattice gas and its surprising behavior}

The KLS model is based on the Ising lattice gas \cite{Ising,YangLee} with
attractive nearest-neighbor (NN) interactions, evolving under particle-hole
or spin-exchange \cite{Kawasaki} dynamics. The original study involved
square lattices with periodic boundary conditions (PBC) \cite{KLS}. Here,
let us consider a slightly more general version, with other BC's on a
rectangular lattice with $L_x\times L_y$ sites (each of which may be
occupied by a particle or left vacant). Thus, a configuration is specified
by the occupation numbers $\{n_{x,y}\}$, where $x,y$ labels a site and $n$
is either 1 or 0. The interparticle attraction is given by the Ising
Hamiltonian: ${\cal H}=-4J\sum n_{x,y}n_{x^{\prime },y^{\prime }}$, where $%
x,y$ and $x^{\prime },y^{\prime }$ are NN sites and $J>0$. With no drive and
coupled to a thermal bath at temperature $T$, a half filled system undergoes
a second order phase transition at the Onsager temperature $%
T_O=(2.2692..)J/k_B$, from a homogeneous, disordered state to an
inhomogeneous state displaying the coexistence of two regions with high and
low particle densities. Minimizing surface energy, each region forms a
single strip, parallel to the shorter axis. To simulate the lattice gas, a
common protocol is to choose a random NN particle-hole pair and exchange
them with probability $\min [1,e^{-\Delta {\cal H}/k_BT}]$, where $\Delta 
{\cal H}$ is the change in ${\cal H}$ due to the exchange.

The deceptively simple extension in KLS is to bias the particle hops along,
say, the $y$ axis, so that the new rates are $\min [1,e^{-(\Delta {\cal H+}%
E\Delta y)/k_BT}]$. Locally, the effect of the ``electric'' field, $E$, is
identical to that due to gravity. However, due to the PBC, this modification
cannot be accommodated by a (single-valued) Hamiltonian. Instead, the system
settles into a {\em non-equilibrium} stationary state with a non-vanishing
global particle current. At first sight, this KLS model appears quite
similar to the Ising case: For $T$ larger than a critical $T_c$, a half
filled system remains in a homogeneous state, while below $T_c$, the system
displays phase segregation. With deeper probing, dramatic differences
surface at {\em all} temperatures. Moreover, many properties are entirely
counterintuitive. Before discussing these surprises, let us point out that
the usual Ising symmetry (particle$\Leftrightarrow $hole, in the lattice gas
language) is violated by the drive, though the combined operation of
particle $\Leftrightarrow $ hole $\oplus $ $y\Leftrightarrow -y$ (known as
CP) is still valid. A closely related model to KLS is the {\em randomly
driven} lattice gas (RDLG), in which the sign of $E$ is chosen randomly,
say, for every hop attempt \cite{RDS}. The effect is similar to a {\em %
two-temperature} Ising lattice gas (TTLG) \cite{KETT}, in which hops along $%
y $ are coupled to another thermal bath at temperature $T^{\prime }$ and
updated with $\min [1,e^{-\Delta {\cal H}/k_BT^{\prime }}]$. For either of
these, the full Ising symmetry is clearly restored. But neither will be
comparable to the equilibrium Ising model and Boltzmann factors will not
describe the stationary distributions of any of the driven systems.

Returning to KLS, the first surprise is how $T_c$ varies as $E$ is
increased. For ``infinite'' $E$, hops aligned with the $y$ axis are accepted
or rejected{\em \ regardless} of $\Delta {\cal H}$. Thus, the drive tends to
break bonds, impair correlations, and increase disorder -- an effect similar
to being coupled to an ``infinite'' temperature bath. Such considerations
would lead naturally to the prediction that the internal energy (average
number of broken bonds) of the system should increase with $E$, so that $T_c$
would decrease. Simulations \cite{KLS} showed quite the opposite: $T_c$
increases, saturating at $\sim 1.4T_O$ for $E=\infty $! Remarkably, this
same shift in $T_c$ is also observed in both the RDLG and the TTLG.
Especially in the TTLG, it is natural to regard our system as coupled to two
baths (with $T^{\prime }>T$). In this sense, we can rephrase the ``first
surprise'' as negative response: The internal energy of our system {\em %
decreases} even though $T^{\prime }$ is {\em increased}. This kind of
``surprising'' behavior and its origins are now reasonably well understood%
\cite{NegResp}, so that arguments can be advanced to predict the class of
drives that would lead to increasing/decreasing $T_c$'s. The general lesson
here is that negative responses can be easily induced in NESM systems.

The next surprise for the early investigators is that, for all $T>T_c$
(where the system is homogeneous), there are long range correlations --
despite the interactions and the dynamics being both short ranged. The
origin of the difference between this behavior and that in the equilibrium
Ising model can be traced to the violation of detailed balance and the
fluctuation dissipation relation (FDR). For a system in $d$ dimensions with
a conserved density, the autocorrelation function is known to decay as $%
t^{-d/2}$. Given the diffusive nature for a non-critical system, the scaling 
$r\sim t^{-1/2}$ should hold and so, we should expect the equal-time
correlation to decay as $r^{-d}$\cite{GG}. The amplitude of this power turns
out to be {\em anisotropic} -- positive along the direction of the drive and
negative otherwise, mimicking a dipolar interaction. In Fourier space, this
amplitude transforms into a {\em discontinuity} singularity\cite{GenLRC} of
the structure factor $S\left( \vec{k}\right) $ at $\vec{k}=0$, e.g., ${\cal %
R }\equiv S\left( k_x\rightarrow 0,k_y=0\right) - S\left(
k_x=0,k_y\rightarrow 0\right) >0$ in $d=2$. This behavior is ``generic,'' in
that ${\cal R}$ is tied to the violation of the FDR. In this sense, systems
in equilibrium are ``singular,'' since FDR forces ${\cal R}$ to vanish, so
that the correlation in the Ising lattice gas decays as $e^{-r/\xi }$ rather
than $r^{-d}$. To understand these features, it is straightforward to
following the spirit of Landau-Ginzburg for the Ising model and formulate a
theory for the coarse grained particle density, $\rho ({\bf x},t)$. Defining 
$\varphi \equiv 2 \rho -1$, taking into account all anisotropies, and
anticipating relevant interactions for $T<T_c$, we write the full Langevin
equation 
\begin{eqnarray}
\partial _t\varphi ({\bf x},t) &=&\lambda \{(\tau _{\bot }-\nabla _{\bot
}^2)\nabla _{\bot }^2\varphi +(\tau _{\Vert }-\alpha _{\parallel }\partial
^2)\partial ^2\varphi -2\alpha _{\times }\partial ^2\nabla _{\bot }^2\varphi
+  \nonumber \\
&&+u(\nabla _{\bot }^2+\kappa \partial ^2)\varphi ^3+{\cal E}\partial
\varphi ^2\}-\xi {\bf (x},t)\,\,,  \label{lang}
\end{eqnarray}
with noise correlations 
\begin{equation}
\langle \xi {\bf (x},t)\xi ({\bf x}^{\prime },t^{\prime })\rangle =-\left(
\sigma _{\bot }\nabla _{\bot }^2+\sigma _{\parallel }\partial ^2\right)
\delta (x-x^{\prime })\delta (t-t^{\prime })\,\,.  \label{eta}
\end{equation}
For the $d=2$ case, $\nabla _{\bot }$ and $\partial $ reduce to $\partial _x$
and $\partial _y$, respectively. This approach can account for all the novel
properties phenomenologically, for example, with ${\cal R}\propto \sigma
_{\bot }/\tau _{\bot }-\sigma _{\Vert }/\tau _{\Vert }$\cite{GenLRC}.

As $T$ is lowered towards $T_c$, simulations first revealed the onset of
phase separation, but only {\em transverse} to the drive, corresponding to a
diverging ${\cal R}$ but with $S\left( k_x=0,k_y\rightarrow 0\right) $
remaining finite. As a result, $k_y$ does not scale naively with the
transverse momenta, so that analyses using techniques for ``strong
anisotropic scaling'' are unavoidable\cite{SAS}. Based on these observations
and starting with the Langevin equation above, a field theoretic
renormalization group analysis can be set up\cite{JSLC} and, unlike the
Ising universality class\cite{Ising class}, the upper critical dimension is $%
d_c=5$. More significantly, the fixed point cannot be written in terms of a
``Hamiltonian'' and is genuinely ``non-equilibrium'' in the sense that it
contains a term corresponding to a non-trivial (probability) current\cite{K*}%
. By contrast, a similar treatment for the RDLG leads to a fixed point {\em %
Hamiltonian}\cite{RDS}, so that its leading singularities fall into the
universality class of a system {\em in equilibrium}. Returning to the KLS
case, a hidden symmetry associated with the fixed point is identified, so
that critical exponents can be calculated {\em to all orders} in $5-d$ \cite
{JSLC}. Thus, we expect these predictions to be reliable down to $d=2$,
without the necessity of Borel resummation\cite{Ising class}. These novel
critical properties are largely confirmed in extensive simulation studies%
\cite{SimCrit}. Despite lingering controversies associated with claims to
the contrary\cite{Marro}, no other field theoretic description is free of
deficiencies at the basic level of symmetries\cite{6a}.

%%%%%%%%%%%%%%%   Fig 1  %%%%%%%%%%%%  
\begin{figure}
%\vspace*{0.7in}
\centerline{\epsfxsize=4.00in\ \epsfbox{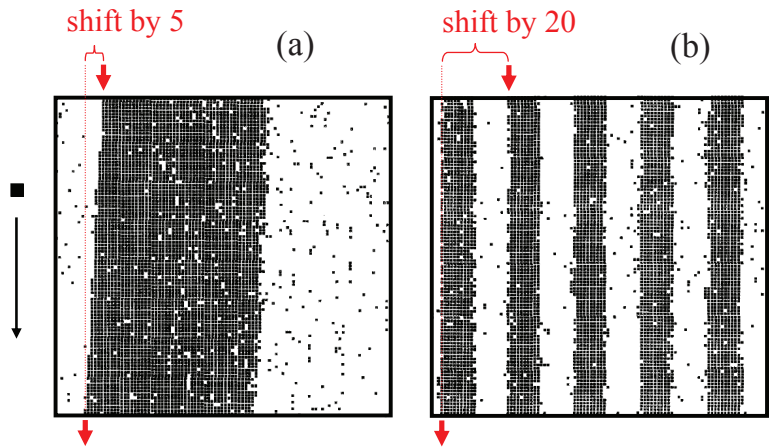}}
\caption{Typical configurations in a $100\times 100$ system, driven
at $E=50J$, with $T=0.8T_O$: 
(a) SPBC imposed with shift 5 and (b) shift 20.}
\end{figure}
%%%%%%%%%%%%%%%%%%%%%%%%%%%%%%%%%%%%%%%
%%%%%%%%%%%%%%%   Fig 2  %%%%%%%%%%%%
\begin{figure}
%%%%%\hspace*{8cm} \epsfxsize=5cm \epsfbox{F2.eps} \vspace*{-1.8cm}
\centerline{\epsfxsize=2.00in\ \epsfbox{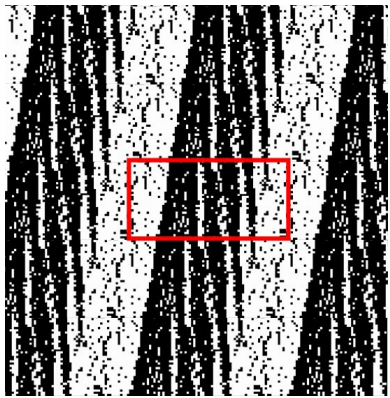}}
\caption{A typical configuration in a $72\times 36$ system with SPBC
and shift $6$, driven at $E=50J$, with $T=0.8T_O$. To provide a global
perspective, the original (framed in a red rectangle) is reproduced 
multiple times in accordance with the SPBC.}
\end{figure}
%%%%%%%%%%%%%%%%%%%%%%%%%%%%%%%%%%%%%%%
%%%%%%%%%%%%%%%   Fig 3  %%%%%%%%%%%%
\begin{figure}
\centerline{\epsfxsize=1.50in\ \epsfbox{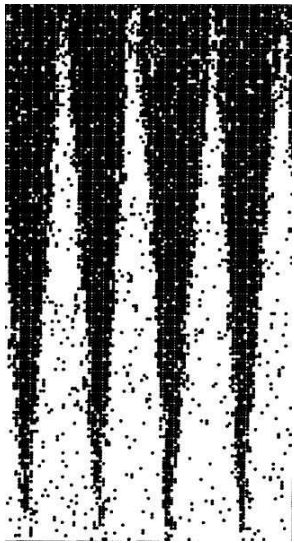}}
\caption{A typical configuration in a $100\times 200$ system with open 
boundaries, driven at $E=2J$, with $T \simeq 0.7T_O$. The top/bottom 
row is filled/emptied at the end of Monte Carlo Step.}
\end{figure}
%%%%%%%%%%%%%%%%%%%%%%%%%%%%%%%%%%%%%%%

Most of the surprising - and poorly understood - phenomena appear for $T<T_c$%
. Simulations showed that the ordered state is similar to the one in the
Ising lattice gas: full phase separation, coexistence of a high-density
region (strip) with a low-density one, interfaces aligned with the drive.
Beyond these gross features in KLS, none are Ising-like and few can be
predicted. Focussing on the steady state first, the most prominent feature
in a phase segregated system is the interface. For $d=2$ systems in
equilibrium, it is always ``rough,'' i.e., it behaves like a random walk,
with a width $w$ that scales as $L^p$, where $p=1/2$. In KLS, simulations
revealed that, especially if the drive is large ($E\gg J$), $p$ is
consistent with zero\cite{L^0}, i.e., the interface is ``smooth.'' Since
interfaces in equilibrium systems can make transitions (the roughening
transition) from being rough to smooth, a natural question is whether
similar transitions exist for the KLS interface. Such an issue has never
been probed systematically. Meanwhile, the roughening transition in
equilibrium is associated with a singularity in the surface tension (i.e.,
interfacial free energy) as function of $\theta $, the angle the normal of
the interface. Whether a similar singularity is present for the KLS
interface motivated a series of Monte Carlo studies\cite{SPBC} of the driven
lattice gas with {\em shifted} periodic boundary conditions (SPBC), a
standard technique for inducing interfaces with various normals. Fig 1(a)
illustrates how ``slanted'' interfaces appear when a shift of 5 is imposed
on a $100\times 100$ lattice. Though there is no well-defined free energy
for a system out-of-equilibrium, the internal energy ($u$) can be easily
measured and a singularity in $\theta $ will also appear in $u\left( \theta
\right) $. The results of these studies led to the next set of surprises
provided by the KLS model. First, not only is $u\left( \theta \right) $
singular ($\partial _\theta u\left( 0\right) $ discontinuous), it is the 
{\em bulk }energy density that is singular (in contrast to the Ising model
with SPBC, where the bulk $u$ is {\em independent }of $\theta $). Further,
as the shift increases, the single ``slanted'' strip breaks up into multiple
strips, as illustrated in Fig 1(b)! Note that the ``multi-strip''
configuration is actually a single strip, with multiple {\em winding} around
the torus. Finally, as the shift is further increased, the system makes a
series of transitions where the winding decreases one by one (``$N$ strips''
merging into ``$N-1$ strips'')! In other words, there appears to be a series
of transitions in the topology of the steady state. Though details remain to
be investigated systematically, many aspects of these transitions have been
explored (through simulations of relatively small systems: $72\times 36$) 
\cite{MJA}. This effort also raised further questions, as more complex
phenomena emerged. For example, an attempt was made at ``catching'' the
system at ``the critical angle'' where the first splitting occurs.
Illustrated in Fig 2, ``icicles'' (long triangular domains) appear on one of
the interfaces. They are ``dynamic,'' continuously growing and shrinking, so
that the system suffers anomalously large fluctuations. From these
explorations, we learn that seemingly trivial modifications of the original
KLS model lead to highly complex and unexpected phenomena. Due to space
limitations, we only list a few others here:

\begin{itemize}
\item  Anomalous correlations in interfaces in both KLS and the RDLG\cite
{KTL-RZ}

\item  Steady states with ``icicles'' in KLS with {\em open} boundary
conditions (illustrated in Fig. 3)\cite{BSZ}

\item  ``Inverted icicles'' during coarsening process and failures of the
continuum theory\cite{ALLZ}

\item  Variety of pattern formation and successes of the continuum theory%
\cite{AR+CY}
\end{itemize}

\noindent Most of these phenomena are far from being well understood. For
example, the aspect ratio of the ``icicle'' pattern seems to be controlled
by the microscopic parameters ($J$,$E$,$T$), but remains to be predicted.

\section{Concluding remarks}

In this article, we presented a very brief summary of the surprises provided
by the original KLS model, i.e., an Ising lattice gas in $d=2$ with uniform
and isotropic {\em attractive interactions}. Thanks to its simplicity,
simulation studies were plentiful and contributed much to the excitement
associated with novel phenomena. Despite its simplicity, however, little is
known analytically, not even in $d=1$ \cite{Bray}. Nevertheless, we have
learned a great deal from it, from specific issues such as generic long
range correlations (induced by a conservation law) to a range of general
properties associated with NESS. Following is a short list of the latter.
(i) Negative responses should not cause a priori alarm, but should be
investigated. (ii) Current loops of probability, mass, energy, etc. --
whether local or global -- can be expected and carry valuable information on
the system. (iii) Familiar routes of the thermodynamic limit are unlikely to
be reliable, while the intuition that provided successful coarse-grained
continuum descriptions may lead us astray. Thus, more dependable techniques
should be developed to arrive at macroscopic properties and mesoscopic
theories. The overall lesson seems to be: Expect the unexpected, whenever
one encounters a new NESS, no matter how trivially it appears to be related
to known systems.

Although there has been only limited progress on the original KLS model,
especially over the last decade, it has spawned considerable activity on
several related fronts. These involve both extensions and simplifications of
the original system, the subject of brief notes in the remaining paragraphs.

Extensions involve anisotropic jump rates\cite{FRL}, anisotropic interactions%
\cite{LBS}, quenched impurities\cite{QI}, multispecies\cite{2sp},
multi-layers\cite{2layers}, mixtures of dynamics\cite{mDyn}, to name just a
few. Remarkably absent are more studies of systems in $d=3$\cite{3d}, in
which new phenomena (e.g., shapes of ``icicles") can be expected. Not
surprisingly, since these models are more complex than KLS, even less is
known analytically (except the fast rate limit\cite{FRL}). Nonetheless,
through computer simulations, these extensions provided many more surprises,
especially when the modifications are so minor that no novel behavior was
anticipacted! Such discoveries further challenge our basic understanding of
NESS: How unpredictably complex phenomena emerge from incredibly simple
dynamic rules\cite{GLife}.

For models {\em simpler} than KLS, on the other hand, enormous advances on
the theoretical front took place. One outstanding case is a lattice gas with 
{\em no} attractive interparticle interaction, i.e., a system of biased
random walkers with on-site exclusion only. The stationary distribution
becomes trivially flat for a system with PBC\cite{FS}, though many
interesting dynamical properties are present\cite{PBCdyn}. With {\em open}
boundary conditions, even the stationary distributions are non-trivial, with
few studies for systems in two or more dimensions. If we simplify further
and consider {\em one-dimensional lattices} (with open boundaries), we find
a wealth of analytic results. Known as the asymmetric exclusion process
(ASEP) and first introduced in 1970\cite{FS}, its exact stationary state was
found\cite{exact}, showing three non-trivial phases as well as unusual
dynamics\cite{dynamics}. Also clarified is its relationship with other
exactly solvable non-equilibrium systems, e.g., the zero range process\cite
{ZRP}. Furthermore, generalizations of ASEP have been exploited to model
physical processes such as protein synthesis (since 1968\cite{MGP}!) and
vehicular traffic\cite{Traffic}. Enjoying considerable attention, it is the
focus of several comprehensive reviews\cite{GS+...}. Of course,
one-dimensional chains cannot support many of the interesting features
discovered in the KLS, e.g., anisotropic correlations, discontinuity
singularities in $S\left( \vec{k}\right) $, and ``icicles." Nevertheless, we
should celebrate these contributions as giant strides towards our
understanding of systems driven far from equilibrium.

To conclude, we remain hopeful that, before long, another Onsager will find
an analytic solution to this amazingly rich, yet simple model, thereby
shedding light on the secrets of NESS in general. Furthermore, our belief is
that, like the Ising model, it will disperse its fruits far afield, beyond
non-equilibrium statistical mechanics to, e.g., graph theory, quantum field
theory, bioscience, neuroscience, sociophysics and econophysics.\\\

{\em Acknowledgements}
In addition to dedicating this article to the 100$^{th}$ Statistical
Mechanics Meeting (where this material was first presented), I thank both
Joel Lebowitz and Herbert Spohn for many, many illuminating discussions. The
support by many collaborators, especially Beate Schmittmann, and the US
National Science Foundation (through DMR-0705152) is also gratefully
acknowledged.


\begin{thebibliography}{99}
\bibitem{mDyn}  There are many systems with mixed dynamics, especially
involving spin-flip, kinetic Ising models as well. See, e.g., Section VII in
ref \cite{DL17}.
\end{thebibliography}
\end{document}